\documentclass[twocolumn,showpacs,amsmath,amssymb]{revtex4}
\usepackage{graphicx}
\usepackage{dcolumn}
\usepackage{bm}
\usepackage{epsfig}
\usepackage{subfigure}

\newcommand{\bi}{\begin{itemize}}
\newcommand{\ei}{\end{itemize}}
\newcommand{\be}{\begin{eqnarray}}
\newcommand{\ee}{\end{eqnarray}}
\newcommand{\bbmatrix}{\left( \begin{array}}
\newcommand{\eematrix}{\end{array} \right)}

\begin{document}

\title{Liquid crystal phases of ultracold dipolar fermions on a lattice}
\author{Chungwei~Lin, Erhai~Zhao\footnote{Present Address: Department of Physics and  Astronomy, George Mason University, 
Fairfax, VA 22030 }, and W.~Vincent~Liu}
\affiliation{Department of Physics and Astronomy, University of Pittsburgh,
Pittsburgh, PA 15260
}
\begin{abstract}
Motivated by the search for quantum liquid crystal phases in a gas of
ultracold atoms and molecules, we study the density wave and nematic
instabilities of dipolar fermions on the two-dimensional square
lattice (in the $x-y$ plane) with dipoles pointing to the $z$
direction. We determine the phase diagram using two complementary
methods, the Hatree-Fock mean field theory and the linear response
analysis of compressibility.  Both give consistent results. In
addition to the staggered ($\pi$, $\pi$) density wave, over a finite
range of densities and hopping parameters, the ground state of the system first becomes
nematic and then smectic, when the dipolar interaction strength is
increased.  Both phases are characterized by the same broken four-fold (C$_4$)
rotational symmetry. The difference is that the nematic phase has a
closed Fermi surface but the smectic does not.  
The transition from the nematic to the smectic phase is associated
with a jump in the nematic order parameter. This jump is closely
related to the van Hove singularities. We derive the kinetic equation
for collective excitations in the normal isotropic phase and find that
the zero sound mode is strongly Landau damped and thus is not a well
defined excitation. Experimental implications of our results are
discussed.

\end{abstract}

\maketitle


\section{Introduction}

It is well known that as the strength of Coulomb interaction is
increased with respect to the kinetic energy, an electron gas goes
from a liquid state to a crystalline phase \cite{Wigner_34}. However
the transition from a liquid to a crystal phase in fermionic systems
with long range interaction may contain several intermediate stages
bearing the name of ``electronic liquid crystal'' phases
~\cite{Fradkin_05}. Analogous to the classical liquid crystals
\cite{Chaikin}, these phases are classified as being ``nematic'' and
``smectic'' according to their symmetry breaking (Fermi surface
deformation) as compared to the isotropic case.  In the nematic phase
\cite{Kivelson_98,Yamase_00-2,Oganesyan_01,Kee_03,Khavkine_04,Quintanilla_06},
the rotational symmetry is broken so the typical Fermi surface has a
cigar-like shape, i.e., it is stretched in one direction and shrunk in
other directions. In the smectic phase the system is effectively in a
reduced dimension \cite{Kivelson_98}, accordingly the Fermi surface is
divided into disconnected pieces. The transition to the smectic phases
is thus naturally connected to dimensional crossover phenomena which
have drawn many interests \cite{Carlson_00,Biermann_01,Ho_04, Kollath_08}.

Electronic nematic order has been observed and studied in a number of solid state
materials, such as transition metal oxides \cite{Tranquada_95,Orenstein_00,Yamase_00-1,Kivelson_04} and
quantum Hall systems (e.g., GaAs/AlGaAs heterostructure in high
magnetic field) \cite{Lilly_99, Pan_99}. These systems are typically
two-dimensional and signatures of nematic order include additional
peaks in neutron scattering \cite{Tranquada_95} and transport
anisotropy \cite{Lilly_99, Borzi_07}.  The nematic order can be viewed  either
as fluctuations (disordering) of static stripe-like ordered
states \cite{Sun_08} or as an instability of the liquid (isotropic) states \cite{Lawler_06}.
Possible nematic order in the two-dimensional Hubbard model has been
extensively discussed in the context of high temperature
superconductors \cite{Tranquada_95,Yamase_00-1,Halboth_00}. 
Away from half filling, a stripe order can be
stabilized by the antiferromagnetic (AF) spin exchange. For example,
at $1/8$ doping, three quarters of sites have one localized electron
with AF spin arrangement maximizing the energy gain from
spin-exchange, while the rest one quarter of sites have an average 0.5
delocalized electron propagating along one particular direction
forming ``conducting veins''; these conducting veins appear every four
lattice constants constituting the stripe phase
\cite{Tranquada_95,Orenstein_00}. The nematic order can thus be viewed
as quantum and/or thermal fluctuations of these static stripes
\cite{Kivelson_98,Lawler_06}. Similar understanding applies to quantum Hall
systems as well \cite{Koulakov_96,Fradkin_99}.

Because of their excellent tunability with dipole moments, cold polar molecular gases have
been proposed as an ideal system to study the electronic liquid
crystal phases
\cite{Mancini_04,Buchler_07,Micheli_07,Sawyer_07,Ni_08}. Under an
external electric or magnetic field, all dipoles are aligned along the
field direction, and the potential energy between two dipoles is
$V(\vec{R})=d^2 [1-3 \cos^2 \theta]/|\vec{R}|^3$, with $d$ the induced
dipole moment, $\vec{R}$ the relative position between the two
dipoles, and $\theta$ the angle between the applied field and
$\vec{R}$. Since the induced dipole moment is proportional to the
external field, by tuning the amplitude and the angle (relative to the
system) of the field one can directly control the strength of
long-range interaction.  Recently, there appeared many theoretical
works on dipolar Fermi gas in the continuum
\cite{Miyakawa_08,Fregoso_09_1,Chan_09,Fregoso_09_2,Baranov_04,  Bruun_08,Zhao_09}. By contrast,
studies on dipolar fermions on lattices are relatively few and focus
on anisotropic lattices \cite{Quintanilla_09}. 

\begin{figure}[htbp]
\vspace{0.02\textwidth}
\begin{center}
   \epsfig{file = 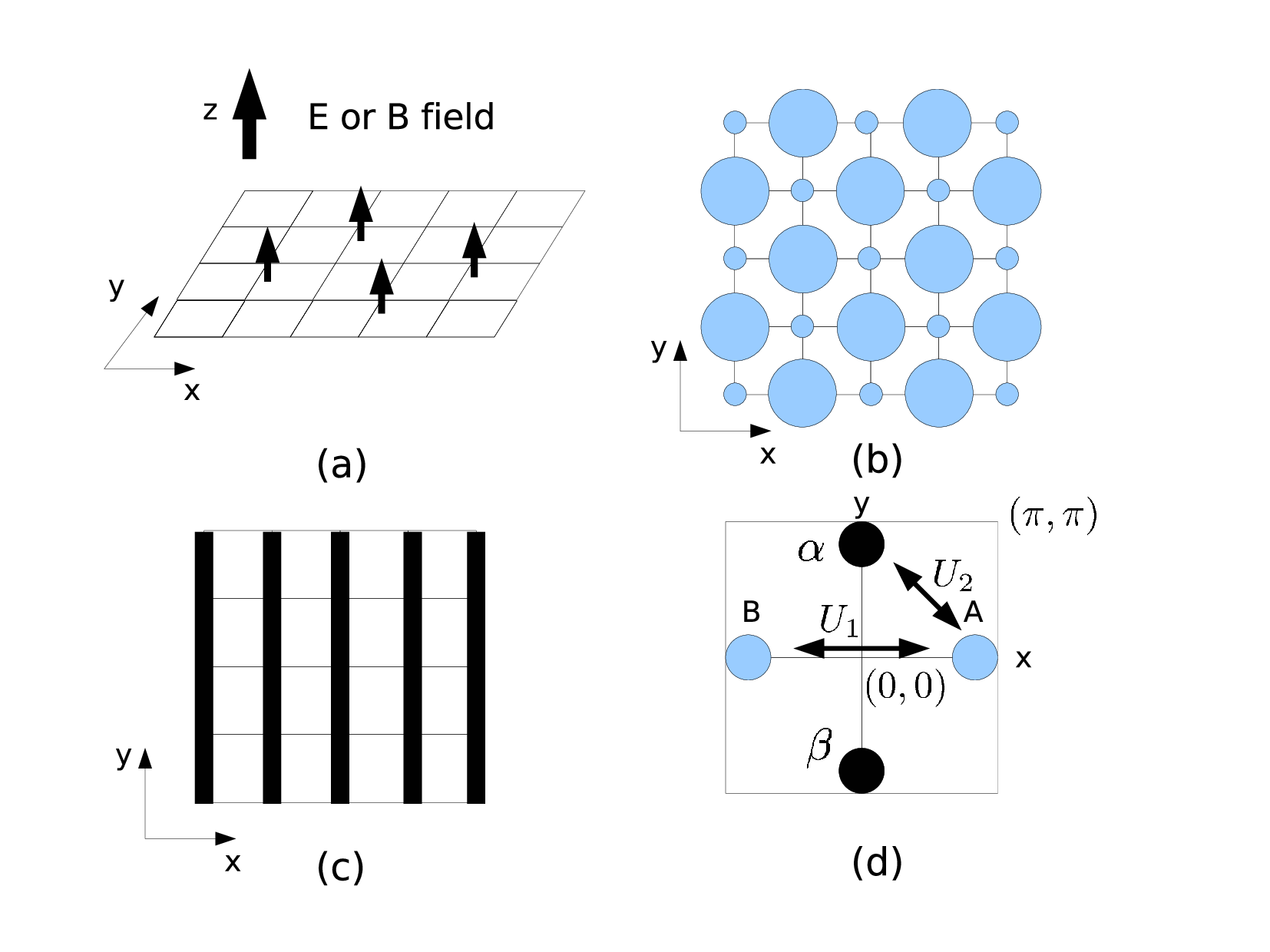,  width = 0.5\textwidth}
   \caption{(a) The experimental setup: a square lattice in the $x-y$ plane and the applied field
	(either electric or magnetic) along the $z$ direction. The dipole moments of fermions are aligned along
	the $z$ direction which leads to large intersite density-density interaction.
	(b) The staggered density wave: the density is more concentrated on one sublattice than the other.
	The size of circle indicates the density at the given site.
	(c) The nematic phase: the effective hoppings along $x$ and $y$ are different.
	(d) The first Brillouin zone of the square lattice. Four circles ($A,B,\alpha,\beta$) mark the regions
	near the van Hove points. The interaction between the opposite van Hove points is $U_1$, and for neighboring van Hove points is $U_2$.
	}
   \label{fig:Phases}
\end{center}
\end{figure}

In this paper we consider the simplest possible system where a single
species of dipolar fermions are loaded into the square optical lattice (in
the $x-y$ plane) \cite{Yamase_05} with the external field along the $z$ direction,
schematically shown in Fig.~\ref{fig:Phases}(a). In this setup the
dipolar interaction has the simple form $d^2/R^3$. We focus on the
instabilities of the normal isotropic phase.   We find that 
the transitions from isotropic to liquid crystal phases are generally of first order.
The transitions to smectic phase are associated with a jump in the order parameter
which is closely related to the van Hove singularities in the low dimensional lattices 
\cite{Halboth_00,Hankevych_02}. 
Our estimate shows that the magnitude of dipole moment 
required to achieve the liquid crystal phases
is within the reach of current experiments of hetero-nuclear polar
molecules. 
The rest of the paper is
organized as follows. In section II, we introduce our model
Hamiltonian and define all relevant phases. We also discuss the van
Hove points in this model and the special features of dipolar
scattering between them.  In Section III, we analyze in detail the
various instabilities from the isotropic state to obtain the phase
diagram of the system. This is done by Hatree-Fock mean field theory
and linear response analysis of the compressibility. 
 Special attention is paid to understand the order of normal-nematic and nematic-smectic transitions.
In section IV we study the collective excitations in the isotropic
phase. We briefly discuss the implications of our results to
experiments in section V before conclude in section VI.

\section{Model and definition}

The general Hamiltonian for single species of dipolar fermions on the square lattice is
\be
H &=&
\sum_{ij} t_{ij} \hat{c}^{\dagger}_{i} \hat{c}_{j}
+ \frac{1}{2}\sum_{ ij } V_{ij} \hat{n}_i \hat{n}_j -\mu \sum_i
\hat{n}_i, \nonumber \\ 
&=& \sum_{\vec{k}} (\epsilon_{\vec{k}} - \mu) \hat{c}^{\dagger}_{\vec{k}} \hat{c}_{\vec{k}}
+\frac{1}{N}\sum_{\vec{k} } V(\vec{k}) \hat{\rho}_{\vec{k}}
\hat{\rho}_{-\vec{k}}. 
\label{eqn:H}
\ee
Here, $t_{ij}$ is the hopping amplitude between site $i$ and $j$,
$V_{ij}\propto d^2/|\vec{r}_i-\vec{r}_j|^3$ is the (repulsive) dipolar interaction between site $i$ and $j$,
$V(\vec{k})$ is the Fourier transform of $V_{ij}$,
$N$ is the total number of sites, $\epsilon_{\vec{k}}$ is the bare (in the absence of $V_{ij}$) band energy dispersion, and $\hat{\rho}_{\vec{k}} = \sum_{\vec{q}} \hat{c}^{\dagger}_{\vec{q}+\vec{k}} \hat{c}_{\vec{q}} $. Since the intersite repulsion takes the form of density-density interaction, this model is sometimes referred to as the extended Hubbard model \cite{Nayak_00}. The actual calculation is done for a given density (particle per site) $n$, and the chemical potential $\mu$ is adjusted to yield the fixed density.

First we give the precise definitions of several phases in our system.
Due to the presence of lattice, the ``isotropic'' or normal phase is a state that has the same
symmetry of the Hamiltonian. In the nematic phase, the C$_4$ rotation symmetry is reduced to C$_2$ but
the lattice translational symmetry still holds in both the $x$ and $y$ direction.
A further constraint is that the Fermi surface is closed.
On the mean field level, the nematic phase can be viewed as the effective hopping amplitudes (renormalized by the dipolar interaction) along
the $x$ and $y$ direction are different, as demonstrated in Fig.~\ref{fig:Phases}(c).
The transition from the isotropic to the nematic phase is also referred to as Pomeranchuk instability
~\cite{Halboth_00, Pomeranchuk_58}. The smectic phase
has the same symmetry as the nematic, but has an open Fermi surface. The transition from the nematic to
the smectic phase is a Lifshitz transition \cite{Abrikosov_88} where the topology of Fermi surface changes. 
Finally, we also consider the possibility of the staggered density wave (sDW) phase, in which
the average density on one sublattice is different from the other sublattice, as illustrated in Fig.~\ref{fig:Phases}(b).

Compared to the continuum gas, an important feature of the two
dimensional lattice is the van Hove singularity in the density of
states. The van Hove points ($\vec{k}_{VH}$) are $\vec{k}$-points in
the reciprocal space with vanishing group velocity
$\vec{\nabla}_{\vec{k}} \epsilon_{\vec{k}} |_{\vec{k}=\vec{k}_{VH} } =
0$.  In two-dimension, this leads to a logarithmic divergence in the
density of states, i.e. the density of states $g(\varepsilon) \sim
-\log |\varepsilon-\varepsilon_{VH}|$ where $\varepsilon_{VH} =
\epsilon_{\vec{k}_{VH}}$ is the van Hove energy.  The lattice symmetry
implies that for a non-zero $\vec{k}_{VH}$, all other $\vec{k}$ points
generated by symmetry transformations of $\vec{k}_{VH}$ are also van
Hove points.  When the chemical potential is close to
$\varepsilon_{VH}$, most low energy excitations are around
$\vec{k}_{VH}$, so interactions between states near the van Hove
points become dominantly important.

Our subsequent discussion will be valid for the general form of the
Hamiltonian Eq.~\eqref{eqn:H}. In the numerical simulation, however,
we choose a specific Hamiltonian as follows.  We keep the first and
second nearest neighbor hopping $t$ and $t'$, which gives
$\epsilon_{\vec{k}} = -2t(\cos k_x+\cos k_y ) - 4t'\cos k_x \cos k_y
$.  Since the strength of dipolar interaction falls off rapidly as a
function of distance ($1/r^3$), we consider the case where the lattice
constant is large enough so that only the nearest neighbor
density-density interaction is kept \cite{Quintanilla_09}.  Under this
simplification the dipolar interaction strength is described by a
single parameter $U$ ($>$0), i.e., in Eq. \eqref{eqn:H} $V_{ij} =
(U/2) \delta_{i,j \pm \hat{a}}$ with $\hat{a} = \hat{x}$ or $\hat{y}$
and $V(\vec{k}) = +U/2 (\cos k_x + \cos k_y )$.  The specific model is
thus parametrized by the hopping $t$ and $t'$, the dipolar interaction
strength $U$, and the filling (density) $n$.  All energies are
measured in unit of $t$ for the remaining discussion.  For this model,
the van Hove points are located at $(0, \pm \pi)$ and $(\pm \pi, 0 )$.
They are labeled by $A,B$ and $\alpha, \beta$, respectively as shown
in Fig.~\ref{fig:Phases}(d).  
 As we shall explicitly show in next section (the discussion below Eq.~(\ref{eqn:Ek_nematic2})),
 $-2V(\vec{k}-\vec{k}')$ is identified as the interaction between states labeled by $\vec{k}$ and $\vec{k}'$.
We point out here that 
the dipolar interaction between opposite
van Hove points (such as $A$ and $B$ in Fig.~\ref{fig:Phases}(d)) is attractive,
\be
U_1\equiv -2V(\vec{k}-\vec{k}')|_{\vec{k}=(-\pi,0),\vec{k}'=(\pi,0)}  = -2U, \label{u1}
\ee
while it is repulsive between neighboring van Hove points (such as $A$ and $\alpha$ in Fig.~\ref{fig:Phases}(d)),
\be
U_2\equiv -2V(\vec{k}-\vec{k}')|_{\vec{k}=(0,\pi),\vec{k}'=(\pi,0)}  = +2U. \label{u2}
\ee
This property of dipolar interaction is very important for
our discussion of the nematic instability in the next section.

\section{Phase diagram}
First, we use Hartree-Fock (HF) approximation to study the ground state of the system Eq. \eqref{eqn:H}
at zero temperature. HF mean field theory has been playing an important role in previous studies of electronic nematic phases
\cite{Kee_03,Quintanilla_06,Quintanilla_09}. 
We consider and compare two possible symmetry breaking phases: the staggered density wave and the nematic phase.
We have also considered the $d$-density wave state \cite{Nayak_00, DDW} but found it has higher energy than the
staggered density wave (also known as $s$-density wave in Ref. \cite{Nayak_00}), so we shall not discuss it in any detail here.

{\em Staggered Density Wave}  (sDW):
In this phase, the fermion density is more concentrated on one of the square sublattices as shown in Fig.~\ref{fig:Phases}(b), so $\langle \rho_{\vec{Q}} \rangle$ is nonzero with $\vec{Q}=(\pi,\pi)$.
This arrangement can reduce the nearest neighbor repulsion energy, which is the dominant interaction energy.
Within the HF approximation, the reduced Hamiltonian for this phase is simply
\begin{align}
H_{HF}^{sDW} = &\sum_{\vec{k}} (\epsilon_{\vec{k}} - \mu) \hat{c}^{\dagger}_{\vec{k}} \hat{c}_{\vec{k}}
+2 V(\vec{Q}) M_{sDW} \hat{\rho}_{\vec{Q}} \nonumber \\ &- N   V(\vec{Q}) M_{sDW}^2.
\label{eqn:H_sDW}
\end{align}
Here, the sDW order parameter $M_{sDW}$ is given by the self-consistent equation $\frac{\partial \langle H_{HF}^{sDW} \rangle}{\partial M_{sDW}} =0$,
\be
M_{sDW} = \frac{1}{N}\sum_{\vec{k}} \langle \hat{c}^{\dagger}_{\vec{k}+\vec{Q}} \hat{c}_{\vec{k}} \rangle.\label{msdw}
\ee
By solving Eq. \eqref{eqn:H_sDW} and \eqref{msdw}, one
can obtain the critical interaction strength $U_c$ above which $M_{sDW}$ becomes nonzero.

{\em Nematic phase}:
In the nematic phase, the system has to rotate $180^{\circ}$, instead of $90^{\circ}$, to
go back to itself. To understand the basic mechanism behind the spontaneous Fermi surface distortion
(which costs kinetic energy), we count the interaction energy between all four van Hove points. As shown in Fig.~\ref{fig:Phases}(d) and discussed above, the interaction between opposite van Hove points
is $U_1$ while between neighboring points is $U_2$. The total interaction energy is
\[
E_{tot} = U_1 (n_A n_B + n_{\alpha} n_{\beta}) + U_2 (n_A+ n_B) (n_{\alpha}+ n_{\beta}).
\]
In the isotropic phase, $n_A = n_B = n_{\alpha} = n_{\beta} = n_0$,
leading to
\[
E^{iso}_{tot} = E_0 = (2 U_1 + 4 U_2) n_0^2.
\]
In the nematic phase, quite generally we have $n_A = n_B = n_0-\delta$,  $n_{\alpha} = n_{\beta} = n_0+\delta$, where $\delta$ characterizes the distortion. This leads to total interaction energy
\[
E^{nem}_{tot} = E_0 + (2U_1-4U_2) \delta^2.
\]
Therefore if $U_1<0$ and $U_2>0$, which we have shown is exactly the case for dipolar interaction in Eq. \eqref{u1} and \eqref{u2}, the nematic phase is energy favored over the isotropic phase
with net energy gain
\[
\Delta E = (2|U_1|+4|U_2|)\delta^2>0.
\]
Similar argument was elaborated by Halboth and Metzner \cite{Halboth_00} in the context of the two-dimensional Hubbard model away from half filling. There, the effective interactions between van Hove points come from a renormalization procedure, while in the present case, the required interactions come directly from the dipolar interaction.

To formulate a HF description of the nematic phase, we notice that because of the lattice translational symmetry, the crystal momentum is still a good quantum number and
the nematic state can be described by a distribution function $n_{\vec{k}} = \Theta(\tilde{\mu} - \tilde{\epsilon}_{\vec{k}})$
\cite{Quintanilla_09,Quintanilla_06}
where $\tilde{\epsilon}_{\vec{k}}$ is the renormalized dispersion to be specified and $\tilde{\mu}$
is the corresponding chemical potential determined by the fermion density.
With this ansatz, the Hartree-Fock Hamiltonian for the nematic phase is
\begin{align}
H_{HF}^{nem} =& \sum_{\vec{k}} (\tilde{\epsilon}_{\vec{k}} - \tilde{\mu}) \hat{c}^{\dagger}_{\vec{k}} \hat{c}_{\vec{k}}
- N   V(0) n^2 \nonumber \\&+\frac{1}{N} \sum_{\vec{k} \vec{k}'} V(\vec{k} - \vec{k}') n_{\vec{k}}  n_{\vec{k}'},
\label{eqn:H_nematic}
\end{align}
with $\tilde{\mu} = \mu - 2 V(0) n$ and
\be
\tilde{\epsilon}_{\vec{k}} (\{n_{\vec{k}} \}) =
\epsilon_{\vec{k}} - \frac{2}{N} \sum_{\vec{k}'} V(\vec{k} - \vec{k}') n_{\vec{k}'}.
\label{eqn:Ek_nematic2}
\ee
Note that $n_{\vec{k}}$ (or equivalently $\tilde{\epsilon}_{\vec{k}}$) has to be solved self-consistently from Eq.~\eqref{eqn:H_nematic} and Eq.~\eqref{eqn:Ek_nematic2}.
From Eq.~\eqref{eqn:H_nematic}, $-2 V(\vec{k} - \vec{k}') $ is identified as the interaction between quasi-particles with momentum $\vec{k}$ and $\vec{k}'$.
For calculations with fixed density, the Hartree term $2 V(0) n$ is independent of $\vec{k}$ and only shifts the chemical potential by a constant, therefore for simplicity we denote the
chemical potential with $\mu$ (instead of $\tilde{\mu}$) in following discussions.
The nematic order parameter $M^{nem}$ can be defined as
\be
M^{nem}=\frac{8}{N}\sum_{0<k_x<k_y<\pi}
( \tilde{ \epsilon}_{k_x,k_y} - \tilde{ \epsilon}_{k_y,k_x} ).
\ee
Note the summation is restricted in the first quadrant of the Brillouin zone and only for $k_x<k_y$, because
parity is conserved in the nematic phase. One notes that the formalism described here
applies to the smectic phase.

\subsection{Phase boundaries}
\begin{figure}[htbp]
\begin{center}
   \epsfig{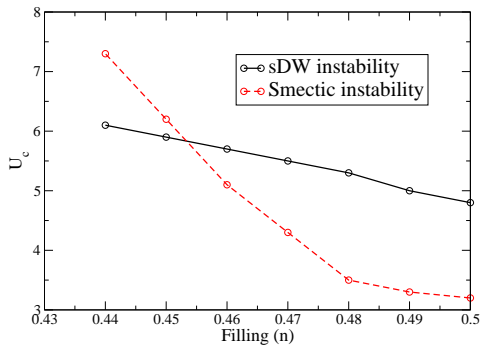}
   \caption{The phase boundaries of staggered density wave (solid) and
   smectic phase (dashed) for $t'=0$, $n=0.44-0.5$. The ordered
   quantum phase is above the respective transition line.} 
   \label{fig:Determine_PB}
\end{center}
\end{figure}

For given $t'$ and $n$, we first compute the critical interaction
strength $U_c$ for both the staggered density wave and the nematic or smectic
phase. Whichever has a smaller $U_c$ is identified as the leading
instability of the system. For example Fig.~\ref{fig:Determine_PB} shows $U_c$
for both phases as a function of $n$ for $t'=0$.  For these parameters
the smectic phase is the leading instability when $0.46<n<0.5$, while
staggered density wave is the leading stability for $n<0.46$. Table I
summarizes the results for several $t'$ and fillings $n$. 
We found when $t'$ is negative enough and the filling is not too close to
the van Hove points, the system first undergoes a weakly $first$ order transition to the
nematic phase before entering the smectic phase. 

\begin{center}
\begin{tabular}{|l||l|l|l|} \hline
 & sDW  & Nematic & Smectic\\ \hline
$t'=0$    & $n<0.46$  & no    & 0.46-0.5\\ \hline
$t'=-0.1$  & $n<0.43$ & no & 0.43-0.47  \\ \hline
$t'=-0.2$  &  $n<0.39$  &  0.39-0.41  & 0.41-0.45  \\ \hline
$t'=-0.3$  &  $n<0.36$   & 0.36-0.38  & 0.38-0.4  \\ \hline
\end{tabular} \\
Table I: The leading instability for $t'=0$, -0.1, -0.2, -0.3 and $n<0.5$.\\
\end{center}

\subsection{The order parameter}
\begin{figure}[htbp]
\begin{center}
   \subfigure[]{\epsfig{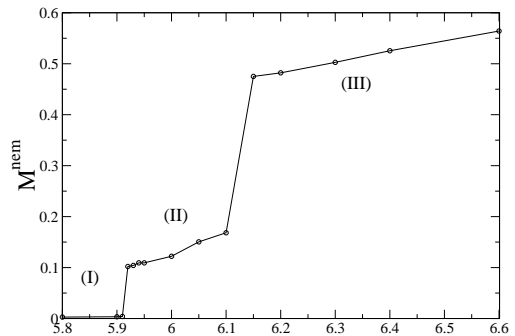}}
   \subfigure[]{\epsfig{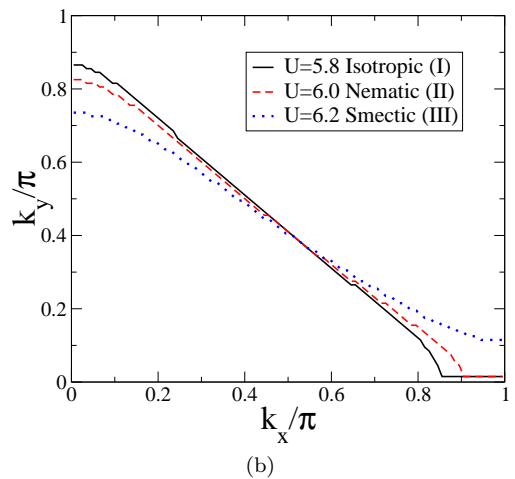}}
   \caption{(a) The nematic order parameter as a function of $U$.  (b)
	The Fermi surface in the first quadrant of the Brillouin zone for $U=5.8$
	(isotropic), 6 (nematic), and 6.2 (smectic).  These results
	are computed at $t'=-0.2$ and $n=0.4$.  }
   \label{fig:Nem_FS}
\end{center}
\end{figure}

Fig.~\ref{fig:Nem_FS}(a) shows the general behavior of nematic order parameter $M^{nem}$ as a function of $U$, 
computed for $t'=-0.2$ and $n=0.4$. The  interaction strength $U$ can be divided into three regions.
For small $U$ (Region I), the system is isotropic as indicated by the Fermi surface for $U=5.8$ (solid curve)
in Fig.~\ref{fig:Nem_FS}(b). When $U>U_c^{nem}\sim 5.9$ (Region II), the symmetry between the $x$ and $y$
direction is broken, the systems enters the nematic phase, and the corresponding Fermi surface, elongated in the 
$x$ direction, is shown (dashed curve) in Fig.~\ref{fig:Nem_FS}(b) for $U=6.0$.
We emphasize that the isotropic-nematic phase transition is of weakly first order. 
Finally, when $U>U_{c2}\sim 6.15$ (Region III), there is another jump in the order parameter corresponding to a Lifshitz transition into the smectic phase. The open Fermi surface for $U=6.2$ is shown (dot line) in Fig.~\ref{fig:Nem_FS}(b). This transition is also referred to as the meta-nematic transition which emphasizes the the jump between non-zero values of the order parameter \cite{Quintanilla_09,Yamase_07}.
We also note that if the filling is such that the Fermi surface is too close to the van Hove points or the 
second nearest neighbor hopping $t'$ is not negative enough (e.g. the first two rows of Table I),
the nematic region II disappears and the system undergoes a first order transition directly 
to the smectic phase. 
In the following we discuss in more details about these two transitions. 

\subsection{The isotropic to nematic transition}

The transition to the liquid crystal phases was shown previously to be first order
for lattice systems \cite{Oganesyan_01,Kee_03}. Following the argument of Ref.~\cite{Oganesyan_01},
one can expand the ground state energy in terms of the order parameter $Q$ as
\be
E(Q)=E(0) + \frac{A}{4} Q^2 + \frac{B}{8} Q^4 + \cdots .
\ee 
The coefficient $B$ is proportional to the cubic correction to the linearized dispersion
around the Fermi momentum which is generally negative for realistic band structures 
\cite{Oganesyan_01,Kee_03}.
For example, for the tight-binding band we consider here 
$\epsilon(\vec{k}) = -2t (\cos k_x+ \cos k_y) - 4t' \cos k_x \cos k_y$,
the cubic term in the expansion of $\epsilon(\vec{k}_f+(q,0)) - \epsilon(\vec{k}_f )$ in $q$
with $\vec{k}_f = (\alpha,0)$ is proportional to $- (t + 2t') \sin \alpha$
which is normally negative ($2 |t'|<t$). Negative $B$ makes the isotropic-nematic transition first order.
 In this case the nematic phase is expected to be stabilized by the higher power term of $Q$ in the energy, for example $Q^6$.

If $\vec{k}_f$ is too close to the van Hove points, the system undergoes phase transition directly from the isotropic to the smectic
phase. On the other hand, if $\vec{k}_f$ is far away enough from van Hove points and
$t'$ is negative enough to reduce the quartic contribution (make $|B|$ smaller), there is
a finite window for stable nematic phase. For the model considered here we found $t'$ has to
be smaller than $-0.2t$ for the nematic phase to occur.

\subsection{The linear response analysis}

To analyze the Fermi surface instability in more detail,
we consider the response $\delta n_{\vec{k}}$ caused by a Fermi surface perturbation $d \mu_{\vec{k}}$ \cite{Nozieres,Lamas_08}.
The perturbation $d \mu_{\vec{k}}$ modifies the effective dispersion
from $\tilde{\epsilon}_{\vec{k}}$ to $\tilde{\epsilon}'_{\vec{k}}$. To the linear order of $d \mu_{\vec{k}}$,
the change $\tilde{\epsilon}'_{\vec{k}} - \tilde{\epsilon}_{\vec{k}} = \gamma d \mu_{\vec{k}}$. Accordingly, \begin{align}
\delta n_{\vec{k}} &= -\delta(\tilde{\epsilon}_{\vec{k}} - \mu)
[\tilde{\epsilon}'_{\vec{k}} - \tilde{\epsilon}_{\vec{k}} -d \mu_{\vec{k}} ] \nonumber \\
&=+ \delta(\tilde{\epsilon}_{\vec{k}} - \mu) \times (1-\gamma  ) d \mu_{\vec{k}}.
\end{align}
We define the momentum-dependent compressibility as $\kappa(\vec{k})=\delta n_{\vec{k}}/d \mu_{\vec{k}} = + \delta(\tilde{\epsilon}_{\vec{k}} - \mu) \times (1-\gamma)$. A stable Fermi surface has
positive compressibility ($1-\gamma >0$).
We have assumed that $\gamma$ is $k$-independent despite the presence
of lattice. The validity of this assumption 
will be established shortly.
Using Eq. \ref{eqn:Ek_nematic2},
one finds
\begin{align}
\tilde{\epsilon}'_{\vec{k}} - \tilde{\epsilon}_{\vec{k}}
= -\frac{2}{N} \sum_{\vec{k}'} V(\vec{k} - \vec{k}')
\delta(\tilde{\epsilon}_{\vec{k}'} - \mu)  (1-\gamma) d \mu_{\vec{k}'}.
\end{align}
This equation, combined with the definition of $\gamma$, leads to the eigenvalue equation 
\be
\sum_{\vec{k}'} c_{\vec{k} \vec{k}'} d \mu_{\vec{k}'}  = \lambda d \mu_{\vec{k}},
\label{eqn:K-Com}
\ee
with $\lambda = \frac{\gamma}{1-\gamma}$ and $c_{\vec{k} \vec{k}'} = -\frac{2}{N}
V(\vec{k} - \vec{k}') \delta(\tilde{\epsilon}_{\vec{k}'} - \mu) $. $1-\gamma > 0$ translates to
$\lambda > -1$, so the condition for stable Fermi surface becomes Det$(c_{\vec{k} \vec{k}'} +1) > 0$.
The delta function in the definition of matrix $c_{\vec{k} \vec{k}'}$ has to be 
treated with care in numerical calculations, this is discussed in the Appendix.

We first discuss the implications and limitations of
Eq.~\eqref{eqn:K-Com}. First, in the eigenvalue equation, the
eigenvector $\{ d \mu_{\vec{k}} \}$ corresponding to the eigenvalue
approaching $\lambda = -1$ provides information about the shape of
Fermi surface in the nematic phase. Second, $c_{\vec{k} \vec{k}'}$
contains a $\delta$  function, indicating that only $k$ points
at the Fermi surface defined by the renormalized
$\tilde{\epsilon}_{\vec{k}}$ are relevant. As explicitly shown in the
appendix, the $\delta$ function further implies that the main
contribution is from $\vec{k}$ points whose renormalized Fermi
velocities $|\vec{ \nabla}\tilde{\epsilon}_{\vec{k}} |$ are smallest,
i.e. where the dispersion is flat and there are plenty of states with
energies close to $\mu$. Those points are related by lattice symmetry
operations (rotations and reflections). When applying
Eq.~\eqref{eqn:K-Com} to determine the instability, the weak $\vec{k}$
dependence of $\gamma$ can be safely ignored. 
Finally, Eq.~\eqref{eqn:K-Com} fails due to divergences
when the Fermi surface crosses the van Hove points.  It also becomes
inapplicable if the phase transition is of first order.

\begin{figure}[htbp]
\begin{center}
   \epsfig{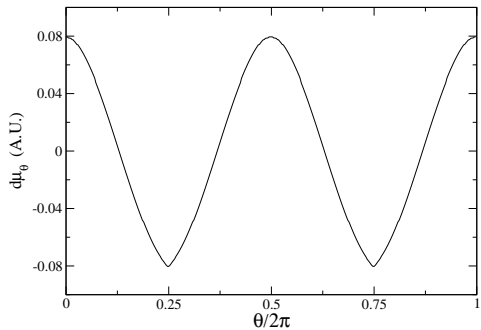}
   \caption{The eigenvector of the softest mode of Eq.~\eqref{eqn:K-Com}
	computed for $U=5.8$, $t'=-0.2$, and $n=0.4$. 
	}
   \label{fig:S_Mode}
\end{center}
\end{figure}

Because the transition is only of weakly first order, 
we  apply Eq.~\eqref{eqn:K-Com} to analyze the transition between
the isotropic (Region I) and the nematic (Region II) phase.   First,
the nematic instability can be detected by the emergence of negative
eigenvalue of matrix $(c_{\vec{k} \vec{k}'}+I)$, where $I$ is the
identity matrix.  We find that for the nematic transition, which we
find is of weakly first order, the critical value $U_c^{nem}$ obtained by
the Hartree-Fock approximation is roughly $5\%$ smaller than
that by analysis using Eq.~\eqref{eqn:K-Com}.  
Actually even for the isotropic-smectic transition, $U_c^{smec}$ from Eq.~\eqref{eqn:K-Com}
is found to be just roughly $10\%$ larger than the mean field result.
Since the Hartree-Fock calculation converges very slowly near the transition, it
is of advantage to determine $U_c$ from the linear response analysis
presented here. Second, the eigenvector of the softest mode, i.e., the
one corresponding to the smallest eigenvalue of $c_{\vec{k}
\vec{k}'}+I$ in Eq.~\eqref{eqn:K-Com}, tells how the Fermi surface
deforms in the nematic phase. Fig.~\ref{fig:S_Mode} shows the softest
eigenmode $d \mu_{\theta}$ for $U=5.8$, $t'=-0.2$, and $n=0.4$, where
the angle $\theta \equiv \tan^{-1}\frac{k_y}{k_x}$ in the polar
coordinate.  Here $d \mu_{\theta}$, the perturbative deformation of the Fermi
surface, is consistent with the Fermi surface in the nematic phase
obtained in HF calculation, the dashed curve in
Fig.~\ref{fig:Nem_FS}(b).

\subsection{The meta-nematic transition}
The order parameter jump across the  nematic-to-smectic or isotropic-to-smectic transition is also closely related to the van Hove singularities.
The Fermi surface is defined by
$\tilde{\epsilon}_{\vec{k}} = \mu$. For a given direction, the change in Fermi momentum  $\triangle k_F$ due to a change in the chemical potential $\triangle \mu$
is proportional to $\frac{\triangle \mu}{| \vec{\nabla}_{\vec{k}} \tilde{\epsilon}| }$.
 Because the area enclosed by the Fermi surface is conserved for a given density
(known as Luttinger's theorem), when
the anisotropy of the nematic phase is increased by increasing $U$, a shrink of Fermi surface in
one direction (say $y$) must be compensated by the expansion in the other (say $x$) direction.
 When the expansion is to include some van Hove points, such as $(\pm \pi,0)$ in Fig.~\ref{fig:Nem_FS}(b), the area increase in that direction
is infinitely large compared to the shrink in the other direction, i.e., $ \triangle k_F (\hat{x}) / \triangle k_F (\hat{y})\rightarrow \infty $.
For this reason the transition from a closed to an open Fermi surface cannot be smooth, which is
reflected on the jump of nematic order parameter. 

\section{Zero sound}
Following the standard Landau Fermi liquid approach \cite{Nozieres}, we derive the quantum kinetic equation to 
determine the collective excitation spectra in the isotropic phase \cite{Nilsson_05, Wolfle_07}. The main question is whether zero sound is 
a well defined collective mode driven by the dipolar interaction.
The starting point is to generalize Eq.~\eqref{eqn:Ek_nematic2} by
assuming slow spatial ($\vec{r}$) and time ($t$) dependence of both the 
distribution function $n_{\vec{k}}(\vec{r})$ and the
effective dispersion $\tilde{\epsilon}_{\vec{k}}(\vec{r})$,
\be
\tilde{\epsilon}_{\vec{k}} (\vec{r}) =
\epsilon_{\vec{k}} - \frac{2}{N} \sum_{\vec{k}'} V(\vec{k} - \vec{k}') n_{\vec{k}'} (\vec{r}),
\label{eqn:Ek_r}
\ee
and assuming quasi-particles are in local equilibrium.
In the collisionless regime, the equation of motion for $\delta n_{\vec{k}}(\vec{r},t)$ is given by
\begin{align}
\frac{\partial}{\partial t}[ \delta n_{\vec{k}}(\vec{r},t)] + \frac{\partial}{\partial \vec{r}} [\delta n_{\vec{k}}(\vec{r},t)]
\frac{\partial \tilde{\epsilon}_{\vec{k}} (\vec{r})}{\partial \vec{k}} + \frac{\partial}{\partial \vec{k}} [\delta n_{\vec{k}}(\vec{r},t)] \nonumber \\
 -\frac{\partial \tilde{\epsilon}_{\vec{k}} (\vec{r})}{\partial \vec{r}}= 0.
\nonumber
\end{align}
Defining $\vec{v}_{\vec{k}} = \vec{\nabla}_{\vec{k}} \tilde{\epsilon}_{\vec{k}}$ and using Eq.~\eqref{eqn:Ek_r},
the above equation becomes
\begin{align}
&\frac{\partial}{\partial t}[ \delta n_{\vec{k}}(\vec{r},t)] + \frac{\partial}{\partial \vec{r}} [\delta n_{\vec{k}}(\vec{r},t)]
\cdot \vec{v}_{\vec{k}}  \nonumber \\
&+ \delta(\tilde{\epsilon}_{\vec{k}}-\mu) \vec{v}_{\vec{k}}\cdot \frac{-2}{N} \sum_{\vec{k}'}
V(\vec{k}-\vec{k}') \frac{\partial [ \delta n_{\vec{k}'}(\vec{r},t) ]}{\partial \vec{r}}
 = 0.
\end{align}
Seeking a wave solution of the form $\delta n_{\vec{k}}(\vec{r},t) =
\delta(\tilde{\epsilon}_{\vec{k}}-\mu) u_{\vec{k}} e^{i (\vec{q}\cdot \vec{r} -\omega_{\vec{q}} t )}$,
we obtain the equation  for zero sound with wave vector $\vec{q}$ and frequency $\omega_{\vec{q}}$ as
\begin{align}
&\sum_{\vec{k}'} D(\vec{q})_{\vec{k} \vec{k}'}  u_{\vec{k}'} = \omega_{\vec{q}} u_{\vec{k}} \label{eqn:0sound} \\
&=\sum_{\vec{k}'} \left[\vec{q} \cdot \vec{v}_{\vec{k}} \delta_{\vec{k} \vec{k}'}
+ \vec{q} \cdot \vec{v}_{\vec{k}} \frac{-2}{N} V(\vec{k}-\vec{k}') \delta(\tilde{\epsilon}_{\vec{k}'}-\mu) \right]  u_{\vec{k}'}. \nonumber
\end{align}
This is again an eigenvalue equation.
Note that both $ D(\vec{q})_{\vec{k} \vec{k}'} $ and $\omega_{\vec{q}}$ are linear in $\vec{q}$.


For our specific model, we find that in the isotropic phase, all
eigenvalues of $D(\vec{q})$ in Eq.~\eqref{eqn:0sound} are real and
bounded from above, $|\omega_{\vec{q}}|<|\vec{q}\cdot \vec{v}_F
|_{max}$. This implies that the zero sound modes overlap with
particle-hole continuum and are strongly Landau damped
\cite{Nozieres}. Therefore zero sound is not an independent, well
defined excitation of the system.  The nematic instability occurs when
the eigenvalue frequency $\omega_{\vec{q}}$ becomes imaginary
(numerically  we find that two of eigenvalues 
become purely imaginary across the nematic transition).  The nematic
boundary determined in this way is also consistent with those obtained
from the Hatree-Fock and linear response analysis.

\section{Experimental implications}
Now we estimate the experimental parameters required to observe the nematic phase using polar molecules.
The optical lattice is characterized by the laser wavelength $\lambda$ (the lattice constant $a_0=\lambda/2$)
and the lattice potential depth $V_0$ is measured in unit of the recoiled energy $E_R = h^2/(2 m \lambda^2)$  \cite{Jordens_08}.
The hopping amplitude is estimated as $t =E_R (2/\sqrt{\pi}) \eta^{3/4} e^{-2 \sqrt{\eta}}$ with $\eta=V_0/E_R$ \cite{Hofstetter_02}.
Typical values of the wavelength is 500-1000 $nm$ and $\eta$ 5-30 \cite{Jordens_08},
leading to $t$ of the order a few or tens of Hertz (multiplied by the Planck constant $h$). 
 For the dipolar interaction energy $U = 2 d^2/ a_0^3$ to reach liquid crystal phases 
(for example, $U \geq 6t$ for the phase diagram shown in Fig.~\ref{fig:Nem_FS}),
 the dipole moment $d$  of a few tenths of Debye is required. For example by taking
$t$ to be 10 Hz and $a_0$ 500 $nm$, $d$ has to be roughly 0.18 Debye such that $d^2/a_0^3 = 60$ Hz.
This value is comparable to those observed in the current experiments \cite{Ni_08}.  
 Finally we mention the anisotropy in the momentum distribution within liquid crystal phases can be directly probed in
the time of flight (TOF) measurements -- after turning off the trap,  the expansions of the dipolar gases in $x$ and $y$ 
directions become significantly different as compared to the isotropic  or sDW phase. 
 According to the local density approximation, the inhomogeneity induced by an external
harmonic potential makes the liquid crystal phase coexist with other phases in the optical lattice. Since normal
and sDW phases both result in isotropic expansions in TOF, the presence of these phases weakens, but cannot eliminate, the
anisotropic signal from the liquid crystal phase.

\section{Conclusion}
We have explored the symmetry breaking phases of single species of
dipolar fermions loaded on the square optical lattice with the
external field perpendicular to the plane.  We find that strong enough
dipolar interaction can drive the system into a nematic and further
into a smectic phase. In particular we find that, apart from the
staggered density wave, for a finite range of filling and hopping the
nematic/smectic phase is the leading instability.  In a simplified picture, the
nematic/smectic instability can be understood as driven by the dipolar
scattering between van Hove points, although one has to bear in mind that
the transition exists even in the absence of van Hove singularities.  
The transition from isotropic to liquid crystal phase is generally of first order.
The transition from the nematic
to the smectic phase is associated with a jump in the nematic order
parameter which is required by Luttinger's theorem and closely related
to Fermi surface passing through the van Hove singularities.
%
%
The zero sound mode in the isotropic phase is found to be strongly
Landau damped and is not a well defined excitation of the system.
Finally, our estimate indicates that the parameter regimes for
the liquid crystal phases are within the reach of experiments in near future. 

\section*{Acknowledgment}
We thank Eduardo Fradkin, Hans Peter B\"uchler, and Han Pu for very helpful discussions. 
This work is supported by
Army Research Office Grant No. W911NF-07-1-0293.


\bibliography{MF_dipolar2}

\begin{thebibliography}{52}
\bibitem[{DDW()}]{DDW}
\bibinfo{note}{Strictly speaking the $d$-density wave is $not$ a mean field
  solution to the extended Hubbard model in the sense that one cannot obtain
  the value of order parameter from a set of self-consistency equations.
  However one could use the variational principle to argue that the $d$-density
  wave has lower energy than the isotropic state. To estimate the critical $U$
  of $d$-density wave, we first decompose the interaction into
  $\frac{-1}{N}\sum_{\vec{k} \vec{k}'} V(\vec{k}-\vec{k}')
  c^{\dagger}_{\vec{k}} c_{\vec{k}+\vec{Q}} c^{\dagger}_{\vec{k}'}
  c_{\vec{k}'-\vec{Q}} $, identifying the coefficient $V'$ common to those in
  $f(\vec{k}) f(\vec{k}')$ with $f(\vec{k}) = \cos k_x - \cos k_y$. We then
  approximate the interaction as $\frac{V'}{N}\sum_{\vec{k} \vec{k}'}
  f(\vec{k}) f(\vec{k}') c^{\dagger}_{\vec{k}} c_{\vec{k}+\vec{Q}}
  c^{\dagger}_{\vec{k}'} c_{\vec{k}'-\vec{Q}} $ where $d$-density wave is
  naturally a mean field solution. Note that by doing so, the so called
  extended $s$-density wave ($f(\vec{k}) = \cos k_x + \cos k_y$) has the same
  critical $U$ as that of $d$-density wave.}
\end{thebibliography}
\appendix
\section{The $\delta$ function}
Here we describe how to numerically implement
Eq.~\eqref{eqn:K-Com}.  Explicitly, we compute 
(replacing $d \mu_{\vec{k}}$ in Eq.~\eqref{eqn:K-Com} by  $u_{\vec{k}}$)
\be
\sum_{\vec{k}'} c_{\vec{k} \vec{k}'} u_{\vec{k}'} = -\sum_{\vec{k}'} \frac{2}{N}
V(\vec{k} - \vec{k}') \delta(\tilde{\epsilon}_{\vec{k}'} - \mu) u_{\vec{k}'}. \nonumber
\ee
In principle, the $\vec{k}'$ summation is over the whole Brillouin zone which makes
$c_{\vec{k} \vec{k}'}$ an $N^2\times N^2$ matrix. However, since the Fermi surface is known,
only $u_{\vec{k}} $ on the Fermi surface are involved in Eq.~\eqref{eqn:K-Com}. To treat
the $\delta$ function properly, we first replace the summation by an integral, i.e.,
\be
\sum_{\vec{k}'} c_{\vec{k} \vec{k}'} u_{\vec{k}'} = \frac{-2}{(2\pi)^2} \int dk'_x dk'_y
V(\vec{k} - \vec{k}') \delta(\tilde{\epsilon}_{\vec{k}'} - \mu) u_{\vec{k}'}, \nonumber
\ee
and then change variable to $\tilde{\epsilon} = \tilde{\epsilon}(k_x, k_y)$ and $\theta = \tan^{-1} \frac{k_y}{k_x}$,
leading to
\be
dk_x dk_y = \left|\frac{\partial k_x}{\partial \tilde{\epsilon} } \frac{\partial k_y}{\partial \theta }
- \frac{\partial k_y}{\partial \tilde{\epsilon} } \frac{\partial k_x}{\partial \theta } \right| d\tilde{\epsilon} d\theta
= 1/|\vec{v}_{\vec{k}} \times \vec{\nabla}_{\vec{k}} \theta| \tilde{\epsilon} d\theta, \nonumber
\ee
where $\vec{\nabla}_{\vec{k}} \theta = (-k_y, k_x)/k^2$ and $\vec{v}_{\vec{k}} = \vec{\nabla}_{\vec{k}} \tilde{\epsilon}$.
Define the Jacobian $J(\vec{k}) = 1/|\vec{v}_{\vec{k}} \times
\vec{\nabla}_{\vec{k}} \theta| $. 
In the $(\tilde{\epsilon}, \theta)$ coordinates, 
the $\delta$ function integration restricts $\vec{k}'$
on the Fermi momentum $\vec{k}_F(\theta')$,  yielding
\be
\sum_{\vec{k}'} c_{\vec{k} \vec{k}'} u_{\vec{k}'} = \frac{-2}{(2 \pi)^2} \int_{0}^{2 \pi} d\theta'
J(\vec{k}_F(\theta')) V(\vec{k}- \vec{k}_F(\theta')) u_{\theta'}. \nonumber
\ee
Then Eq.~\eqref{eqn:K-Com} becomes a discretized equation of
\be
\lambda u_{\theta}= \frac{-2}{(2 \pi)^2} \int_{0}^{2 \pi} d\theta'
J(\vec{k}_F(\theta')) V(\vec{k}_F(\theta)- \vec{k}_F(\theta')) u_{\theta'}.\nonumber
\ee
One notices that due to the Jacobian,  the $\vec{k}$ points of
  smaller Fermi velocities contribute more to the 
integral.  However, when $\vec{v}_{\vec{k}}=0$, the Jacobian diverges
and the above integral is not well defined anymore 
(logarithmically divergent).

\section{Erratum: Liquid crystal phases of ultracold dipolar fermions on a lattice}
Due to a numerical error in solving equation (4) and (5) of Ref. [\onlinecite{cLin_10}], the
region where the staggered density wave (sDW) is stabilized was underestimated.
Figure 2 and Table I in Ref. [\onlinecite{cLin_10}] are incorrect. After the correction, we find
that for all fillings, sDW is the leading instability for the specific model in Ref. 
[\onlinecite{cLin_10}] with $t=1$, $t'\in(-0.3,0)$.

\begin{figure}[htbp]
\subfigure[]{
   \epsfig{file = 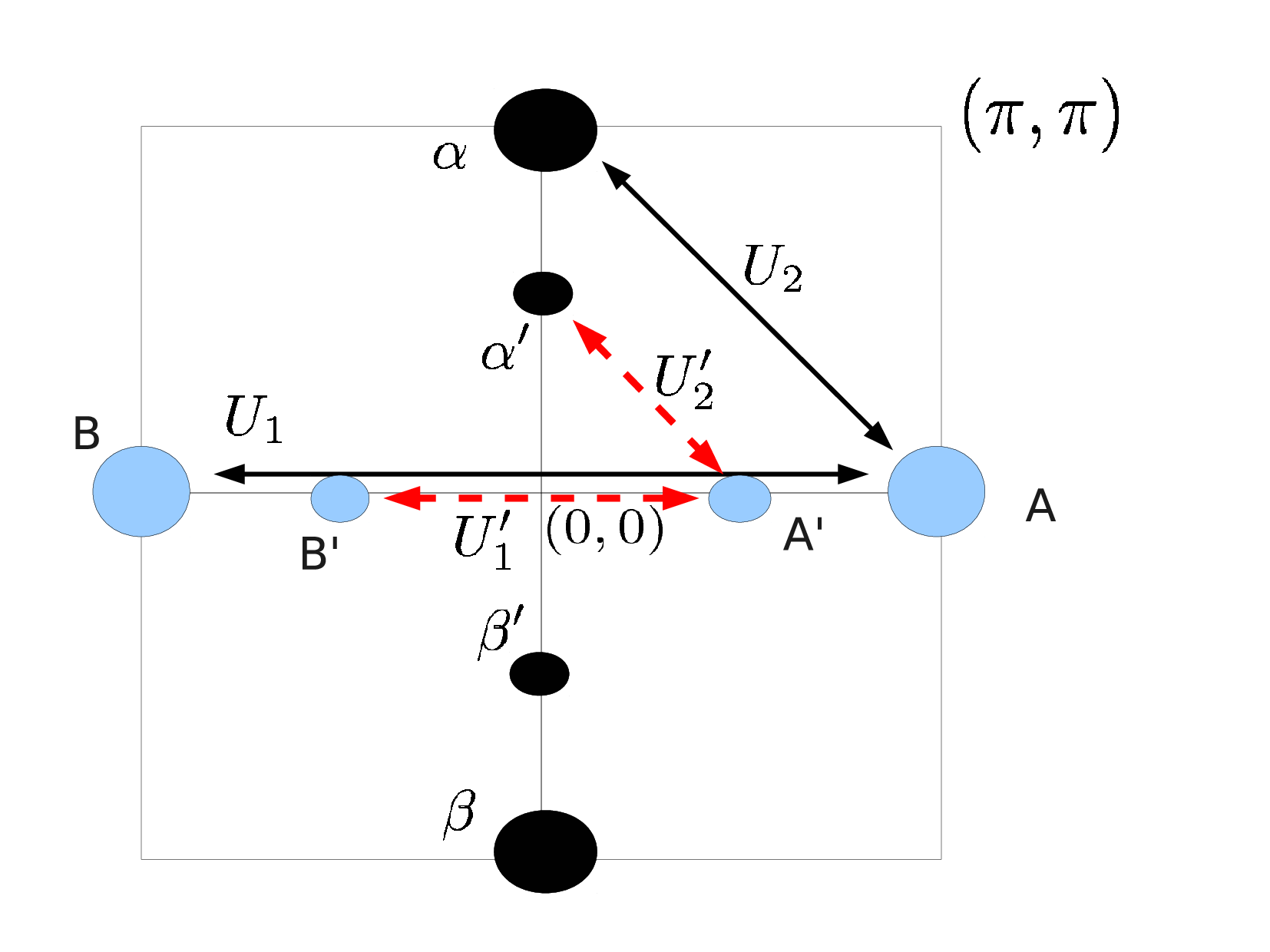,  width = 0.23\textwidth}}
\subfigure[]{   
   \epsfig{file = Uc_n_t2n_-0.4_t3n_0.2-2.eps,  width = 0.23\textwidth}}
   \caption{
   (a) Four new VH points at  $(\pm \bar{k}, 0)$ ($A'$ and $B'$)
  and $(0, \pm \bar{k})$ ($\alpha'$ and $\beta'$) emerge when $t''$ (3rd nearest neighbor hopping) is included.
 (b) The instability boundaries of staggered density wave (solid), 
   liquid crystal (dashed) for $t'=-0.4$, $t''=0.2$ and $n=0.1-0.3$. 
  }   \label{fig:VH_points}
\end{figure}

Here we consider a slightly more general model
\be
H = \sum_{\vec{k}} (\epsilon_{\vec{k}} - \mu) \hat{c}^{\dagger}_{\vec{k}} \hat{c}_{\vec{k}}
+\frac{1}{N}\sum_{\vec{k} } V(\vec{k}) \hat{\rho}_{\vec{k}}
\hat{\rho}_{-\vec{k}}
\label{eqn:H_A}
\ee
with $\epsilon_{\vec{k}} = -2t(\cos k_x+\cos k_y ) - 4t'\cos k_x \cos k_y - 2t''(\cos 2 k_x+\cos 2 k_y )$ 
and $V(\vec{k}) = U/2 (\cos k_x + \cos k_y )$. The introduction of $t''$,
namely the 3rd nearest neighbor hopping, modifies the bare band structure ($U=0$) to
give rise to a new set of van Hove (VH) points as shown in Fig. \ref{fig:VH_points}(a). We now show that liquid crystal (LC) phase 
occurs in this model. The underlying mechanism is the same as outlined in Ref. \cite{cLin_10}:
the effective interaction between neighboring VH points is repulsive,
$U_2'>0$, while that between opposite VH points is attractive, $U_1'<0$, thus giving
energy incentive for breaking the $C_4$ rotational symmetry.

Fig. \ref{fig:VH_points}(b) shows the sDW and liquid crystal instabilities for $t=1$, $t'=-0.4$, $t''=0.2$.
For fillings $0.15<n<0.24$, $U_c^{LC}<U_c^{sDW}$ so the liquid crystal phase will be realized 
as the dipolar interaction is increased.
We have checked that $U_c^{LC}$ reaches minimum when the Fermi surface crosses the new set of VH points.
Similarly for $t=1$, $t'=-0.4$, $t''=0.15$, we find the liquid crystal is the leading instability
for fillings $0.167<n<0.24$.

\begin{figure}[htbp]
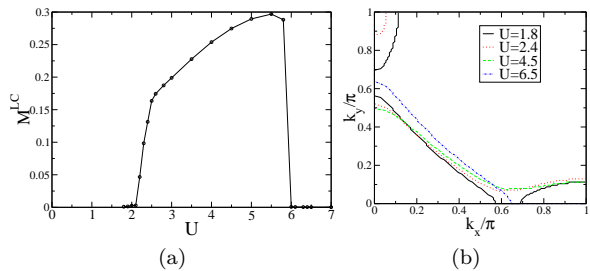

   \subfigure[]{\epsfig{file = Order_U_t1_-0.4_t2_0.2_n0.2.eps,  height = 0.17\textwidth}}
   \subfigure[]{\epsfig{file = FS2_t1.0_t2n-0.40_t3n0.20_U1.8-6.5_T0.010_occu0.20_Kpoints120.eps,  height = 0.17\textwidth}}
   \caption{(a) The liquid crystal order parameter as a function of $U$ for $t=1$, $t'=-0.4$, $t''=0.2$ at $n=0.2$.
	(b) The Fermi surfaces in the first quadrant of the Brillouin zone for $U=1.8$ (isotropic), 
	2.4 and 4.5 (liquid crystal), and 6.5 (isotropic).
	 }
   \label{fig:Order_vs_U_Re}
\end{figure}

Fig. \ref{fig:Order_vs_U_Re} shows the liquid crystal order parameter as a function
of $U$ and the representative Fermi surfaces at different $U$ values, for $t=1$, 
$t'=-0.4$, $t''=0.2$ at $n=0.2$. 
At small $U$, the isotropic (normal) phase
contains five particle-pockets centered at $(0,0)$, $(\pm \pi,0)$, and $(0,\pm\pi)$. 
The topology of Fermi surface within the liquid crystal phase changes for $U$ around 2.5.
Finally, for $U>6$, the isotropic phase reappears with a single 
Fermi surface centered at $(0,0)$. This transition is of first order.

These features can be understood by a careful analysis of the 
renormalization of the dispersion by the dipolar interaction
within the Hartree-Fock mean field theory of \cite{cLin_10}.
Especially, the energy landscape evolves differently near the 
two sets of VH points as $U$ is increased, leading to the nontrivial
Fermi surface evolution in Fig. \ref{fig:Order_vs_U_Re}(b).

In conclusion, liquid crystal phases can occur in two dimensional dipolar systems.
Lattice systems with a band structure 
containing two sets of VH points are particularly promising to develop the
liquid crystal instability. 
The physical mechanism is the same as 
outlined in Ref. \cite{cLin_10}. While the phase boundary presented 
in Ref. \cite{cLin_10}
was wrong, the discussions on the compressibility and
zero sound remain valid. We thank Jim Freericks for drawing our attention to
the error in Ref. \cite{cLin_10}.

\end{document}